\documentclass[preprint,12pt]{elsarticle}

\usepackage{amssymb}
\usepackage{amsmath}
\usepackage{newunicodechar}
\newunicodechar{≥}{\geq}
\usepackage{multirow}
\usepackage{graphicx}
\usepackage[table]{xcolor}
\usepackage{booktabs}
\usepackage{makecell}
\usepackage{algorithm}
\usepackage{algpseudocode}
\usepackage{float}
\journal{Nuclear Physics B}

\begin{document}

\begin{frontmatter}


\title{Slide-Level Active Learning Reduces Annotation Burden in H\&E images}
% Adaptive Patient-Level Active Learning for Annotation-Efficient Histopathology


\title{}


%% Authors
\author[inst1,inst2,inst4]{Mahsa Vali\corref{cor1}}
\ead{mahsavali14@gmail.com}

\author[inst5]{Zhilong Weng}

\author[inst1,inst2] {Noémie Moreau}

\author[inst5]{Yuri Tolkach}

\author[inst1,inst2,inst3]{Katarzyna Bozek}

%% Corresponding author
\cortext[cor1]{Corresponding author}

%% Affiliations
\affiliation[inst1]{
    organization={Institute for Biomedical Informatics, Faculty of Medicine and University Hospital Cologne, University of Cologne},
    city={Cologne},
    country={Germany}
}

\affiliation[inst2]{
    organization={Center for Molecular Medicine Cologne (CMMC), Faculty of Medicine and University Hospital Cologne, University of Cologne},
    city={Cologne},
    country={Germany}
}

\affiliation[inst3]{
    organization={Cologne Excellence Cluster on Cellular Stress Responses in Aging-Associated Diseases (CECAD), University of Cologne},
    city={Cologne},
    country={Germany}
}

\affiliation[inst4]{
    organization={Faculty of Mathematics and Natural Sciences, University of Cologne},
    city={Cologne},
    country={Germany}
}

\affiliation[inst5]{
    organization={Institute of Pathology, University Hospital Cologne},
    city={Cologne},
    country={Germany}
}

% \end{frontmatter}

% \affiliation{organization={},%Department and Organization

%% Abstract
\begin{abstract}
Deep learning-based segmentation of histopathology whole-slide images (WSIs)
requires large amounts of pixel-level annotations, which are costly and
time-consuming to obtain. Active learning (AL) has been proposed to reduce
this effort, but existing methods exhibit three key limitations. Uncertainty
estimation is unreliable on partially annotated WSIs, patch-level acquisition
is inconsistent with slide-level annotation workflows, and class imbalance in
multi-class settings is not explicitly addressed. To address these challenges,
propose SHAL (Slide-level Hybrid Active Learning), a patient-level AL framework, is introduced for annotation-efficient multi-class histopathology segmentation. SHAL integrates three complementary components. A foreground-aware strategy suppresses bias from unlabeled background regions. A stage-adaptive mechanism hybridizes predictive entropy and epistemic uncertainty across learning stages. A class-aware strategy prioritizes diagnostically relevant tissue classes.
SHAL is evaluated on the TCGA colorectal cancer dataset. It achieves the highest
Macro Dice at the full annotation budget (0.846) and reaches Dice~$\geq$~0.80
using only 26\% of the budget (50 of 190 slides), whereas competing methods
reach this threshold only at 37\% (70 slides). Across five independent external
cohorts, SHAL attains the highest mean external Macro Dice~FG (0.815) and the
smallest internal-to-external generalization gap among all methods ($0.025$ at
Round~3, $0.026$ at the full budget). The results indicate that patient-level
hybrid uncertainty acquisition reduces annotation cost without sacrificing
cross-domain generalization in computational pathology.
\end{abstract}

\end{frontmatter}



\section{Introduction}
\label{sec1}

Computational pathology increasingly integrates artificial intelligence (AI)
into digital pathology workflows, enabling quantitative analysis of tissue
morphology and supporting clinical decision-making~\cite{zhang2025artificial,
tan2025computational}. A central task in this area is semantic segmentation,
which provides pixel-level delineation of tissue components such as tumor,
stroma, necrosis, lymphocytes, and other structures. Accurate characterization
of these regions is essential for understanding the tumor microenvironment,
assessing disease progression, and developing robust prognostic
biomarkers~\cite{lobanova2024artificial}. However, producing dense pixel-level
annotations for whole-slide images (WSIs) remains a major bottleneck in
computational pathology~\cite{campanella2019clinical}. WSIs are extremely large,
often reaching gigapixel resolution, and their annotation requires extensive
work by expert pathologists~\cite{nan2025deep}. This process is time-consuming,
expensive, and subject to inter-observer variability~\cite{montezuma2025annotation}.
Unlike natural image datasets, annotation in histopathology requires specialized
expertise and cannot be easily crowdsourced~\cite{shi2023ebhi}. As a result,
annotation cost significantly limits the scalability of AI-based pathology systems.

Deep learning has substantially advanced histopathology segmentation.
Convolutional encoder–decoder architectures, such as U-Net and its variants,
have been widely adopted due to their effectiveness in biomedical image
segmentation~\cite{ronneberger2015u, graham2019hover}. More recently, Transformer
and hybrid architectures have improved the modeling of long-range dependencies
and complex tissue structures, achieving state-of-the-art performance in medical
image analysis~\cite{chen2021transunet, hatamizadeh2022unetr, li2024deep}.
Despite these advances, such models remain highly dependent on large-scale
pixel-level annotations, motivating strategies that reduce annotation cost.

Active learning (AL) has been widely used as a data-efficient strategy to
reduce annotation burden by iteratively selecting informative unlabeled samples
for expert annotation~\cite{settles2009active}. In medical image segmentation,
AL is applied alongside convolutional, Transformer-based, or hybrid architectures
to improve annotation efficiency without modifying the underlying segmentation
model~\cite{BUDD2021102062, WANG2024103201}.

Despite the promise of AL for medical image segmentation, existing AL methods
for histopathology segmentation face several important challenges. Most existing
approaches rely on patch-level or region-level acquisition
strategies~\cite{9648200, 10.1007/978-3-031-43895-0_9}, which often leads to
redundant acquisition of highly similar tissue regions and does not fully align
with slide-level pathology workflows~\cite{campanella2019clinical}. Recent studies
have explored WSI-level AL strategies for histopathology
analysis~\cite{KANG2024102455}. However, slide-level acquisition strategies for
multi-class histopathology segmentation remain largely underexplored. In addition,
many AL approaches rely on a single uncertainty measure for sample acquisition.
Different uncertainty metrics capture complementary aspects of model uncertainty,
and relying on a single acquisition criterion may lead to suboptimal sample
selection in complex multi-class histopathology segmentation
tasks~\cite{NEURIPS2019_95323660, GAILLOCHET2023102958}. A further limitation is
the severe class imbalance commonly observed in histopathology datasets, where
diagnostically important tissue classes may occupy only a small fraction of the
slide area~\cite{schachtsiek2023cbda, YEUNG2022102026}. As a result, acquisition
strategies that do not account for class distribution tend to underrepresent
these minority classes, limiting segmentation performance on clinically relevant
regions.

A complementary line of work reduces annotation effort through semi-supervised
learning (SSL), leveraging large collections of unlabeled
WSIs~\cite{weng2025similarity}. While such methods improve pseudo-label quality
and representation learning, they mainly optimize model training \emph{after}
data selection and typically assume access to large unlabeled datasets. In
contrast, we address the complementary problem of annotation-efficient sample
acquisition \emph{before} annotation, within a patient-level AL framework.

In this study, we approach the selection of data for labeling as a slide-level
AL problem for multi-class histopathology segmentation. Our goal is to achieve
competitive segmentation performance with significantly fewer annotated patients
while aligning the acquisition process with real-world clinical annotation
workflows. To address the challenges above, we introduce SHAL (Slide-level
Hybrid Active Learning), a hybrid uncertainty-based acquisition framework designed
for annotation-efficient histopathology segmentation.
First, we introduce a foreground-aware uncertainty estimation strategy to
suppress uncertainty contributions from unlabeled background regions in partially
annotated WSIs. Specifically, uncertainty computation is limited to annotated
tissue regions to reduce acquisition bias arising from extensive non-annotated
areas.
Second, we propose a stage-adaptive uncertainty mechanism that dynamically
integrates predictive entropy and epistemic uncertainty, estimated via Monte
Carlo dropout~\cite{gal2016dropout, vali2025active}. In early AL iterations, the
framework prioritizes epistemic uncertainty to select globally informative
samples under limited supervision. In later stages, greater emphasis is placed
on predictive entropy to refine decision boundaries around challenging tissue
regions.
Third, we introduce a targeted acquisition bonus to increase sampling of
underrepresented tissue classes, thereby mitigating class imbalance in the
acquired training set. Patch-level acquisition scores are adjusted using the
predicted probability mass of selected minority tissue classes, promoting the
selection of diagnostically relevant but rare tissue regions. Slide-level scores
are computed by aggregating patch-level uncertainty estimates through a
top-$k$ selection strategy, thereby aligning patch-level scoring with slide-level
annotation workflows.

\section{ Materials and methods}

\subsection{Dataset}
\label{sec:dataset}

For model development, we utilized hematoxylin and eosin (H\&E) stained WSIs from the colorectal cancer cohort of The Cancer Genome Atlas (TCGA), as described in~\cite{weng2025similarity}. The dataset consists of 245 WSIs with expert-generated pixel-level annotations covering multiple histological tissue classes, including tumor, stroma, benign epithelium, lymphocytes, necrosis, mucosa, debris, and adipose tissue. The cohort was randomly divided at the patient level into training, validation, and internal test sets to avoid data leakage between patients. Specifically, 190 WSIs were used as the AL pool for model development and iterative sample acquisition, while 30 WSIs were reserved for validation, and 25 WSIs were held out as an internal test set for final evaluation. AL experiments were conducted exclusively on the training pool, whereas the validation and test sets remained fixed throughout all experiments to ensure a fair comparison between acquisition strategies.

To evaluate out-of-distribution generalization performance, we incorporated five independent external cohorts (WNS: Wiener Neustadt, Austria; LMU: Ludwig Maximilian University of Munich, Germany; Charité: Charité University Hospital Berlin, Germany; KAMEDA: Kameda Medical Center, Japan; UKK: University Hospital Cologne, Germany; and the CRAG dataset) exclusively for external validation \cite{weng2025similarity}. The external evaluation set consisted of 70 WSIs collected from five international pathology centers (9 WNS, 10 LMU, 10 Charité, 11 KAMEDA, and 30 UKK slides), in addition to 210 annotated image patches from the CRAG benchmark dataset. These datasets originate from multiple pathology centers and include substantial variability in staining protocols, scanner characteristics, and acquisition conditions. This multi-center variability provides a rigorous out-of-distribution evaluation setting for assessing the robustness and generalization capability of the proposed method.

\subsection{Overview of the Proposed Framework}

\begin{figure*}[t]
    \centering
    \includegraphics[width=\textwidth]{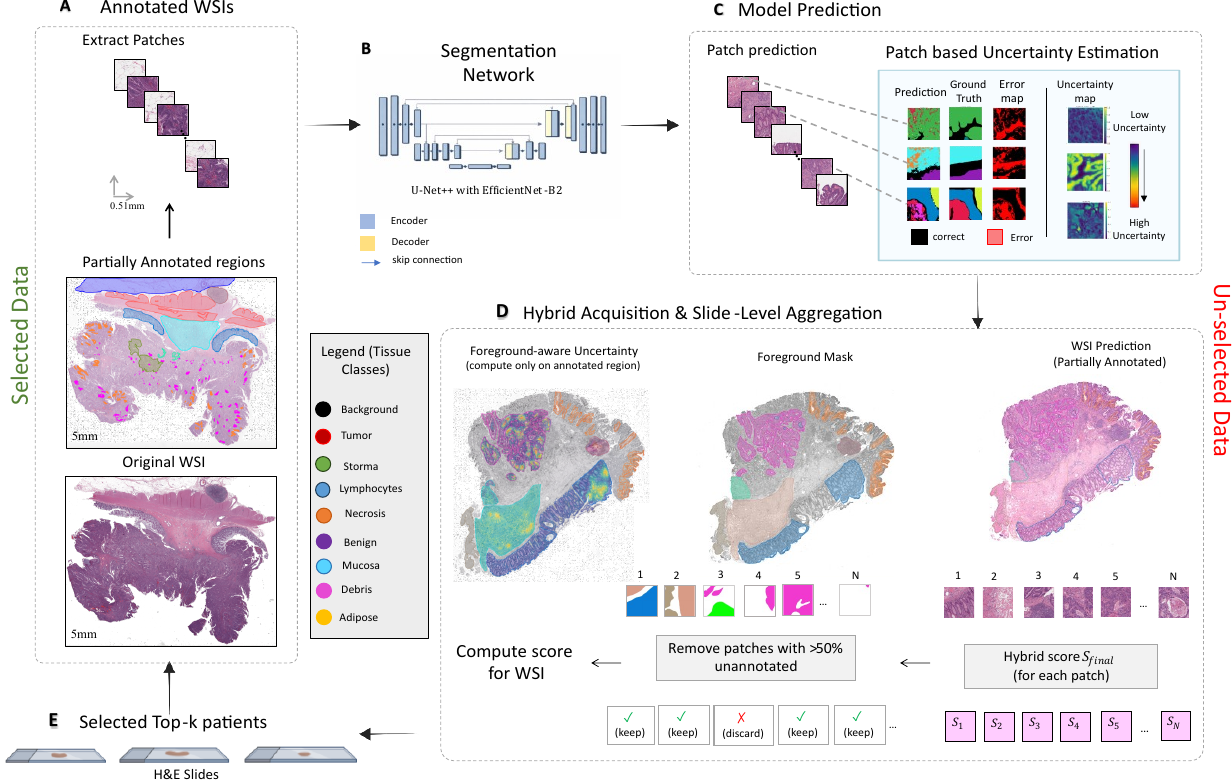}
    \caption{
Overview of the proposed SHAL framework for patient-level AL in histopathology segmentation. Starting from a set of partially annotated WSIs, image patches are extracted from the selected training slides (A) and used to train the segmentation network (B). The trained model is then applied to unlabeled WSIs to generate segmentation predictions and estimate patch-level uncertainty (C). A hybrid acquisition strategy combines foreground-aware uncertainty estimation with target class-aware weighting to compute an acquisition score for each patch while excluding unreliable regions (D). Finally, patch scores are aggregated into slide-level acquisition scores to rank candidate WSIs, and the top-$k$ most informative patients are selected for annotation in the next AL iteration (E).}
    \label{fig:pipeline}
\end{figure*}

In this study we focus on multi-class semantic segmentation of histopathology WSIs under limited annotation budgets, defined as the total number of patient slides selected for training during the AL process. The goal is to achieve performance comparable to fully supervised learning while minimizing the number of slides used for training. All WSIs were digitized at high resolution (typically 40$\times$, approximately 0.25 micrometers per pixel) using clinical-grade scanners. For computational efficiency, each WSI was partitioned into image patches of size $512 \times 512$ at 10$\times$ magnification. Due to the dense annotation protocol, only selected tissue regions within each WSI were annotated at pixel-level precision, while large portions of the slides remained unlabeled and were treated as background during training (see Section~\ref{sec:foreground} for details). Each patch is associated with pixel-wise annotations across $C$ tissue classes.
To reduce the amount of required manual annotation, we adopt a pool-based AL framework~\cite{settles2012active} that iteratively selects the most informative slides for training. Unlike conventional AL approaches that operate at the patch level, our framework performs acquisition at the  patient (slide) level. This design better reflects clinical annotation workflows, where pathologists typically review and annotate slides belonging to a patient rather than isolated patches. 
It also allows the model to capture inter-patient morphological variability, promoting a more diverse and representative training set.
Let $\mathcal{D}_S$ denote the selected training set and $\mathcal{D}_U$ the pool of labeled but unselected slides. At each AL round, the segmentation model is trained on $\mathcal{D}_S$, and an acquisition function evaluates samples from $\mathcal{D}_U$. The highest ranked patients are transferred to $\mathcal{D}_S$, progressively expanding the training set. This iterative process gradually expands the training set while minimizing the total number of slides required to reach competitive segmentation performance. The overall framework is illustrated in 
Figure~\ref{fig:pipeline}.

\subsection{Segmentation Network}

For the segmentation backbone, we adopt the U-Net++ architecture~\cite{zhou2018unetplusplus}, which extends the original U-Net through nested and dense skip connections that reduce the semantic gap between encoder and decoder feature maps. The encoder is based on EfficientNet-B2~\cite{tan2019efficientnet}, pretrained on ImageNet~\cite{deng2009imagenet}, providing multi-scale feature representations suitable for complex histopathological structures~\cite{weng2024grandqc}. The network produces pixel-wise predictions for all tissue classes. To improve segmentation performance on imbalanced and small tissue regions commonly observed in histopathology images, the segmentation model is trained using a combination of Dice loss~\cite{sudre2017generalised} and Focal loss~\cite{lin2017focal}. Dice loss encourages overlap-based optimization for imbalanced segmentation regions, while Focal loss emphasizes difficult and underrepresented pixels during training.

\subsection{Hybrid Uncertainty Acquisition}
\label{sec:hybrid}
Standard uncertainty-based acquisition strategies rely on a single measure applied uniformly across all patches and rounds~\cite{settles2012active}. In the context of multi-class histopathology segmentation with partial annotations, this approach has several important limitations. First, it conflates predictive uncertainty from informative tissue regions with uncertainty arising from unannotated background areas~\cite{weng2025similarity}. Second, it applies a fixed uncertainty measure regardless of the current training stage of the model~\cite{nath2020diminishing}. Third, it treats all tissue classes equally in highly imbalanced datasets~\cite{folmsbee2022histology, cui2019class}. To address these limitations, we propose a hybrid acquisition function composed of three complementary components, described in the following subsections.

\subsubsection{Foreground-Aware Uncertainty Estimation}
\label{sec:foreground}

Uncertainty sampling is one of the most widely used acquisition strategies in AL \cite{settles2012active}. In segmentation tasks, predictive entropy derived from the predicted class probability distribution is commonly used to quantify uncertainty. In partially annotated WSIs, only selected tissue regions are provided with pixel-level annotations, while substantial portions of the tissue area remain unlabeled and are therefore treated as background during training. As a result, patches may contain meaningful tissue structures that are incorrectly considered as background, introducing a systematic bias in uncertainty estimation. Conventional entropy-based strategies may therefore assign high uncertainty scores to such regions, even though they do not provide a useful supervisory signal. To address this limitation, we introduce a foreground-aware uncertainty estimation strategy. For each image patch, a pixel-wise uncertainty map $U \in \mathbb{R}^{H \times W}$ is computed from the model predictions. In the case of entropy-based acquisition, the uncertainty map is estimated using predictive entropy:

\begin{equation}
U(h,w) = - \sum_{c=1}^{C} p_c(h,w)\log p_c(h,w),
\end{equation}
where $p_c(h,w)$ denotes the predicted probability for class $c$ at pixel $(h,w)$.

Let $p_{\text{bg}}(h,w)$ denote the predicted background probability at pixel $(h,w)$, we define a foreground weighting term as:
\begin{equation}
w_{\text{fg}}(h,w) = 1 - p_{\text{bg}}(h,w).
\end{equation}

The foreground-aware patch-level uncertainty score is then computed as a spatially weighted mean:
\begin{equation}
U_{\text{fg}} = \frac{1}{HW} \sum_{h,w} 
w_{\text{fg}}(h,w) \cdot U(h,w),
\end{equation}

where $H$ and $W$ denote the dimensions of the patch. The foreground-aware score $U_{\mathrm{fg}}$ serves as the entropy-based term of
the hybrid acquisition function (Section~\ref{sec:hybrid}). It is combined there
with epistemic and class-aware components, and the resulting patch scores are
aggregated at the WSI level to rank candidate slides.

% This formulation suppresses uncertainty contributions from background dominated and unannotated regions, while emphasizing patches containing diagnostically relevant tissue structures with reliable annotations. As a result, the acquisition process focuses on informative foreground regions that provide a meaningful supervisory signal.

\subsubsection{Stage-Adaptive Combination}

In conventional uncertainty-based AL, a single acquisition function is typically used throughout all rounds. However, the effectiveness of different uncertainty (aleatoric and epistemic uncertainty) measures depends on the training stage of the model. In early rounds, the model is undertrained and exhibits high predictive ambiguity, making entropy-based acquisition effective for identifying informative samples. As training progresses and the model becomes more confident, entropy alone becomes less informative, and model uncertainty becomes more relevant. This can be effectively captured using Monte Carlo (MC) dropout~\cite{gal2016dropout}.

% To account for this behavior, we propose a stage-adaptive acquisition strategy that combines entropy-based and MC dropout-based uncertainty (estimated using the BALD criterion~\cite{houlsby2011bald, gal2016dropout}). This design is motivated by prior observations that the effectiveness of uncertainty measures evolves during the training process~\cite{nath2021diminishing}. The patch-level acquisition score at round $r$ is defined as:

% \begin{equation}
% S^{(r)} = (1-\lambda(r)) \cdot H_{\text{fg}} + \lambda(r) \cdot U_{\text{BALD,fg}},
% \end{equation}

% where $H_{\text{fg}}$ represents the foreground-aware predictive entropy score defined in Equation~(3).  $U_{\text{BALD,fg}}$ is the foreground-aware epistemic uncertainty estimated via MC dropout. It is computed analogously by replacing $U(h,w)$ in Equation~(3) with the BALD uncertainty map:

% \begin{equation}
% U_{\text{BALD}}(h,w) = H[y \mid x] - \mathbb{E}_{p(\omega)}\left[H[y \mid x, \omega]\right],
% \end{equation}

% where $H[y \mid x]$ denotes the predictive entropy over $T$ stochastic forward passes, and $\mathbb{E}_{p(\omega)}[H[y \mid x, \omega]]$ represents the expected entropy under the dropout distribution, capturing epistemic uncertainty.

To account for this behavior, we propose a stage-adaptive acquisition strategy that combines entropy-based and MC dropout-based uncertainty, estimated using the BALD criterion \cite{houlsby2011bayesian, gal2017deep}.
This design is motivated by prior observations that the effectiveness of uncertainty measures evolves during training \cite{nath2020diminishing}:
predictive entropy is informative when the model is undertrained, whereas epistemic (model) uncertainty becomes more discriminative as the model converges. Both terms are obtained by applying the same foreground-aware aggregation operator from Equation~(3) to two different pixel-wise uncertainty maps. The patch-level acquisition score at round $r$ is
defined as:

\begin{equation}
S^{(r)} = (1-\lambda(r)) \cdot U_{\text{ent,fg}}
        + \lambda(r) \cdot U_{\text{BALD,fg}},
\end{equation}

where $U_{\text{ent,fg}}$ is the foreground-aware predictive entropy obtained by aggregating the entropy map of Equation~(1) via Equation~(3), and $U_{\text{BALD,fg}}$ is the foreground-aware epistemic uncertainty obtained by aggregating the BALD map defined below through the same operator.

The BALD map quantifies the mutual information between the prediction and the model parameters, and is estimated via MC dropout with $T$ stochastic forward passes~\cite{houlsby2011bayesian, gal2017deep}. Let
$\hat{p}^{t}(h,w)$ denote the predicted class distribution at pixel $(h,w)$ on the $t$-th forward pass. The BALD uncertainty map is computed as:

\begin{equation}
U_{\text{BALD}}(h,w) =
\underbrace{H\!\left[\tfrac{1}{T}\textstyle\sum_{t=1}^{T} \hat{p}^{t}(h,w)\right]}_{H[y \mid x]}
-
\underbrace{\tfrac{1}{T}\textstyle\sum_{t=1}^{T} H\!\left[\hat{p}^{t}(h,w)\right]}_{\mathbb{E}_{p(\omega)}\left[H[y \mid x, \omega]\right]},
\end{equation}

where $H[\cdot]$ denotes the Shannon entropy over the $C$ tissue classes. The first term is the predictive entropy of the mean prediction, while the second is the expected entropy across passes; their difference isolates epistemic uncertainty from aleatoric noise.

% where both entropy and BALD-based uncertainty are computed in a foreground-aware manner as described in the previous subsection.

The weighting function $\lambda(r)$ controls the contribution of each term and follows a linear decay:

\begin{equation}
\lambda(r) = \frac{r-1}{R-1},
\end{equation}

where R is the number of AL rounds and which satisfies $\lambda(1)=0$ and $\lambda(R)=1$. This results in a smooth transition from entropy-based exploration in early rounds to epistemic uncertainty in later stages, without introducing additional hyperparameters.

% \begin{equation}
% \lambda(r) = 1 - \frac{r-1}{R-1},
% \end{equation}

% which satisfies $\lambda(1)=1$ and $\lambda(R)=0$. This results in a smooth transition from entropy-based exploration in early rounds to dropout-based uncertainty in later stages, without introducing additional hyperparameters.

\subsubsection{Targeted Class-Aware Acquisition}

Certain tissue classes appear infrequently in the dataset and are more challenging to segment accurately. To improve model performance on such underrepresented classes, we introduce a targeted acquisition strategy that prioritizes samples likely to contain these structures. During acquisition, the uncertainty-based score of each patch is scaled by a class-aware weighting term. Specifically, the final acquisition score at round $r$ is defined as:

% \begin{equation}
% S_{\text{final}}^{(r)} = S_{\text{unc}}^{(r)} \cdot \left(1 + \alpha \cdot f_{\text{target}}\right),
% \end{equation}

% where $S_{\text{unc}}^{(r)}$ denotes the stage-adaptive uncertainty score defined in the previous subsection, $f_{\text{target}}$ represents the predicted probability mass of pixels belonging to the target classes, and $\alpha$ is a hyperparameter controlling the strength of the class-aware weighting.

\begin{equation}
S_{\text{final}}^{(r)} = S^{(r)} \cdot \left(1 + \alpha \cdot f_{\text{target}}\right),
\end{equation}

where $S^{(r)}$ denotes the stage-adaptive uncertainty score defined in Equation~(4), $f_{\text{target}}$ represents the predicted probability mass of pixels belonging to the target classes, and $\alpha$ is a hyperparameter controlling the strength of the class-aware weighting.

This formulation biases the acquisition process toward samples that are both uncertain and likely to contain rare or diagnostically relevant tissue patterns. Importantly, the class-aware term acts as a multiplicative scaling factor and does not modify the underlying uncertainty estimation. As a result, the model is encouraged to focus on informative regions associated with difficult and underrepresented classes, improving performance in imbalanced settings.
%patterns.

\subsection{Slide-Level Score Aggregation}

A single WSI may contain thousands of patches, many of which exhibit high redundancy. Selecting slides directly at the patch level may therefore lead to redundant acquisitions, where similar tissue regions are repeatedly selected \cite{folmsbee2022histology, settles2012active}. To address this issue, patch-level acquisition scores are aggregated into a single patient-level score using a top-$K$ mean strategy.

Specifically, for each patient in $\mathcal{D}_U$, all extracted patches are first scored using the hybrid acquisition function $S_{\text{final}}^{(r)}$. Patches dominated by background regions (i.e., containing more than $50\%$ background pixels in the ground truth annotation) are discarded. The filtered patches are then ranked in descending order based on their acquisition scores, and the patient-level score is computed as the mean of the top-$K$ highest-scoring patches:

% \begin{equation}
% \text{Score}_{\text{patient}} =
% \frac{1}{K} \sum_{k=1}^{K} S_{\text{final}}^{(k)},
% \end{equation}

\begin{equation}
\text{Score}_{\text{patient}} =
\frac{1}{K} \sum_{k=1}^{K} S_{\text{final}}^{(r)}(p_k),
\end{equation}

where $K = 20$ and $p_k$ refers to the $k$-th highest-scoring patch of the  patient at round $r$. This strategy ensures that the final score reflects  the most informative regions rather than being dominated by less relevant  or background-heavy patches. Finally, patients are ranked according to their  aggregated scores, and the top-ranked candidates are transferred from  $\mathcal{D}_U$ to $\mathcal{D}_S$ for the next training round. This aggregation process corresponds to the patch ranking and top-$K$ selection illustrated in Figure~\ref{fig:pipeline}(E).

\section{Experimental Setup}

\subsection{Implementation Details}

Following the pool-based AL paradigm~\cite{settles2012active}, SHAL is initialized with a seed set of 10 randomly selected patients. At each AL round, 20 patients are transferred from $\mathcal{D}_U$ to $\mathcal{D}_S$, and the segmentation model is retrained on the expanded selected set using a warm-start strategy initialized from the best checkpoint of the previous round. The annotation budget is defined as the total number of patient slides selected for training during the active learning process. The iterative acquisition procedure continues until this budget is exhausted. To account for changes in class distribution as the selected training set evolves across AL rounds, class weights are dynamically updated using the effective number of samples formulation proposed in~\cite{cui2019class}.

We compare the proposed SHAL method against four baseline acquisition strategies representing different AL paradigms: (i) Random Sampling, which selects slides uniformly at random; (ii) Entropy Sampling, based on predictive entropy; (iii) MC Dropout/BALD~\cite{gal2016dropout}, which estimates epistemic uncertainty using stochastic forward passes; and (iv) CoreSet~\cite{sener2018active}, a diversity-based acquisition strategy that selects representative samples in the feature space. These baselines were selected to represent complementary AL paradigms, including uncertainty-based and diversity-based acquisition strategies commonly used in medical image analysis~\cite{ren2021survey, wang2024comprehensive}. All methods share the same segmentation backbone and training protocol to ensure a fair comparison between acquisition strategies. Performance is evaluated using Dice score, and pixel accuracy on a fixed held-out internal test set. Thresholds used for the annotation-efficiency analysis were evaluated on the validation set, while final performance results are reported on the independent test set.

\section{Results}

\begin{figure}[H]
\centering
\includegraphics[ width=1\textwidth]{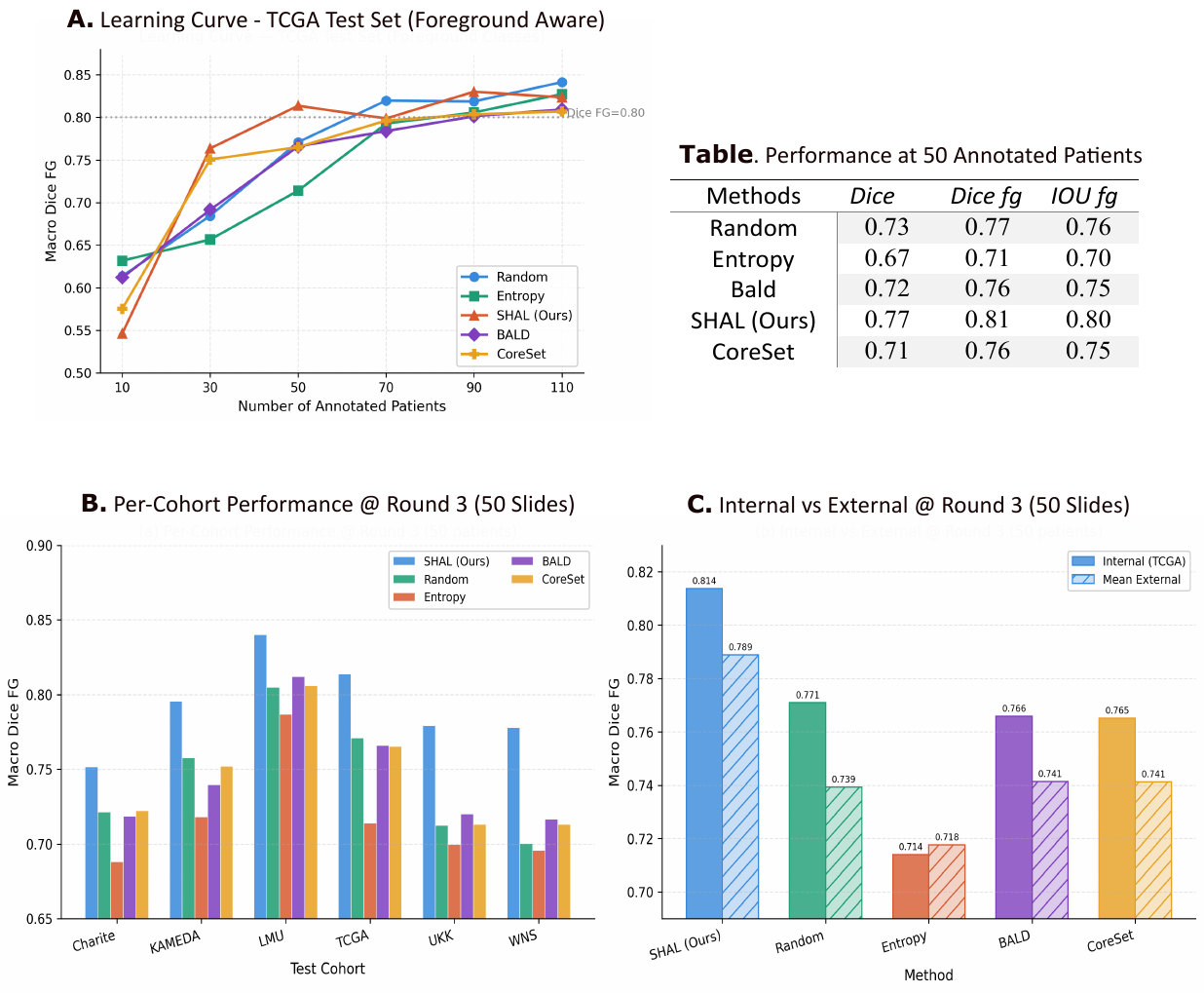}

\caption{
Comprehensive evaluation of AL acquisition strategies across six rounds 
(10--110 annotated patients) and at an annotation budget of 50 patients. \textbf{(A)} Learning curve on the TCGA test set (foreground classes only) across active learning rounds. SHAL achieves competitive performance with fewer annotated patients and reaches Dice FG $\geq$ 0.80 at 50 annotated patients. \textbf{Table.} Segmentation performance on the TCGA test set at 50 annotated patients. SHAL achieves the highest Dice (0.77), Dice FG (0.81), and IoU FG (0.80), demonstrating improved annotation efficiency compared with competing acquisition strategies. \textbf{(B)} Per-cohort Macro Dice FG performance at Round 3 (50 annotated patients) across the internal TCGA cohort and five external cohorts (Charité, KAMEDA, LMU, UKK, and WNS). \textbf{(C)} Comparison of internal (TCGA) and mean external cohort performance at Round 3. SHAL achieves the highest internal and external Dice FG scores while maintaining strong cross-cohort generalization under a limited annotation budget.
}
\label{fig:performance_vs_slides}
\end{figure}

We first evaluate the effectiveness of different acquisition strategies in reducing annotation requirements for multi-class histopathology segmentation. As shown in Figure~\ref{fig:performance_vs_slides}(A), all methods exhibit rapid performance improvement during the early AL rounds, with the largest gains observed between Round~1 and Round~3. SHAL achieves the highest final Macro Dice of 0.846 at Round~6 and remains among the top-performing acquisition strategies throughout the AL process. Entropy-based sampling performs competitively in early rounds but plateaus in later stages, while BALD shows slower initial improvement before converging toward the performance of the other uncertainty-based strategies. CoreSet consistently underperforms compared to uncertainty-based strategies across all rounds. Figure~\ref{fig:performance_vs_slides}(B) summarizes annotation efficiency as a function of the number of annotated patients. SHAL reaches Macro Dice~$\geq$~0.80 at Round~3 (50 patients), whereas all competing methods require at least 70 annotated patients to reach the same threshold, as summarized in Table~\ref{tab:annotation_efficiency}.

\begin{figure}[t]
    \centering
    \includegraphics[
        width=\textwidth,
        trim=0 1.5cm 0 1.5cm,
        clip
    ]{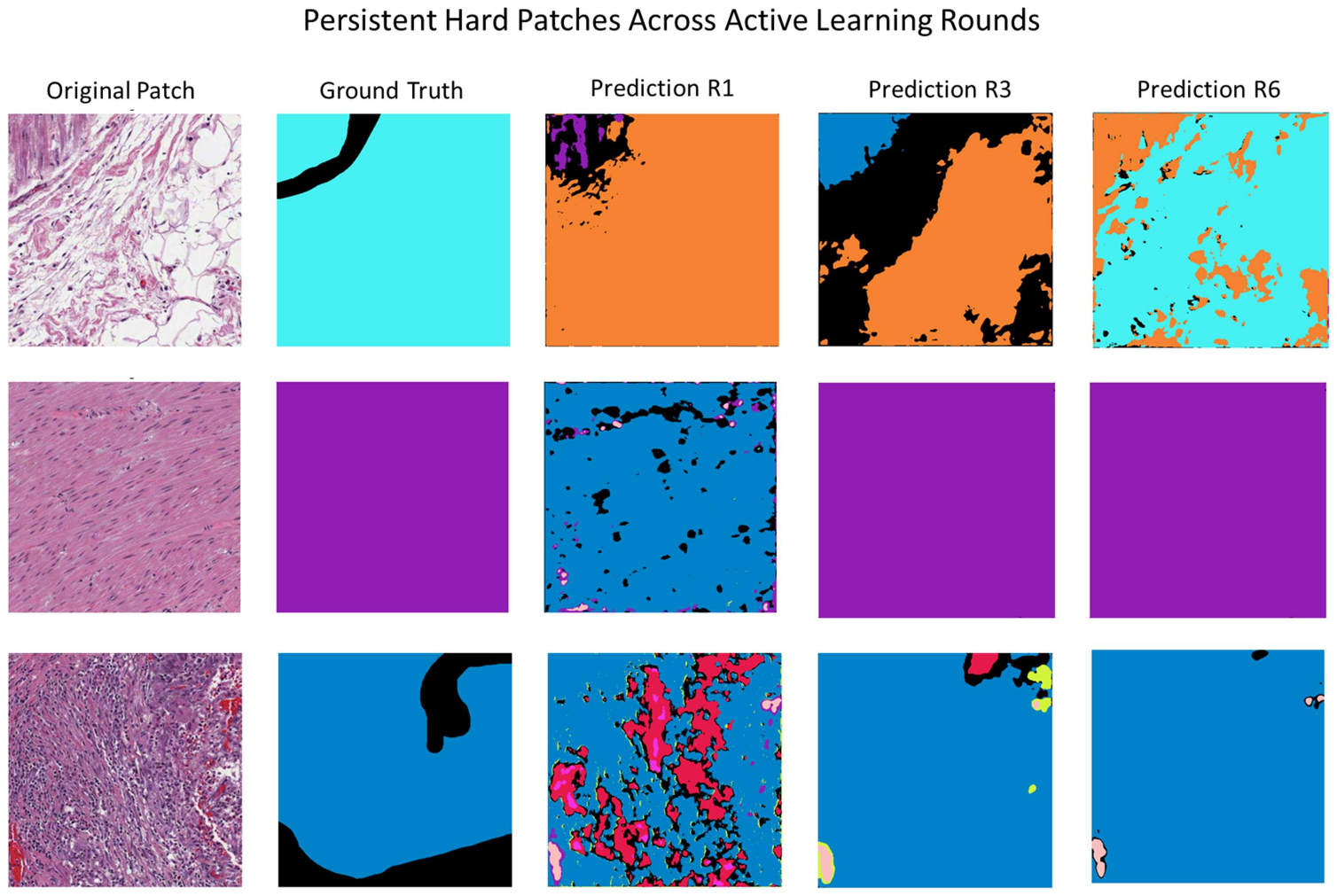}
    \caption{Representative examples of persistently challenging tissue patches. }
    \label{fig:qualitative_results}
\end{figure}

% ###################################################،[ِTODOOOOOOOOOOOOOOOOOOOOOOOOOOOOOOOOOOOOOOOO
% Figure~\ref{fig:qualitative_results} presents qualitative segmentation results along with corresponding uncertainty maps across several representative patches. 
% The first three columns show the input image, ground truth, and model prediction, respectively, along with the Dice score for each sample. The error maps highlight regions where the model fails to accurately capture class boundaries, particularly in complex tissue structures.

% Entropy-based uncertainty provides a coarse indication of ambiguous regions; however, it often lacks specificity and tends to respond to general texture variability. 
% In contrast, BALD better captures epistemic uncertainty, focusing on regions where the model exhibits disagreement under stochastic forward passes.

% The proposed hybrid acquisition strategy further refines uncertainty estimation by incorporating both foreground-aware weighting and target class relevance. 
% As observed, the hybrid maps concentrate on diagnostically meaningful structures and boundary regions, which are more informative for improving model performance during active learning.

% These results confirm that the proposed method not only improves segmentation accuracy but also provides more reliable and interpretable uncertainty estimates compared to baseline approaches.

Figure~\ref{fig:qualitative_results} illustrates representative patches selected by SHAL from a WSI together with the corresponding uncertainty maps and acquisition components. High-ranking patches are typically associated with morphologically complex tissue regions, tissue interfaces, and diagnostically relevant structures exhibiting elevated predictive entropy and epistemic uncertainty. The foreground-aware uncertainty maps suppress background-dominated regions, while the target-class probability highlights regions containing clinically relevant tissue classes. These examples provide qualitative insight into how SHAL prioritizes informative regions during slide acquisition for annotation-efficient segmentation.

\begin{figure}[t]
    \centering
    \includegraphics[ width=1\textwidth]{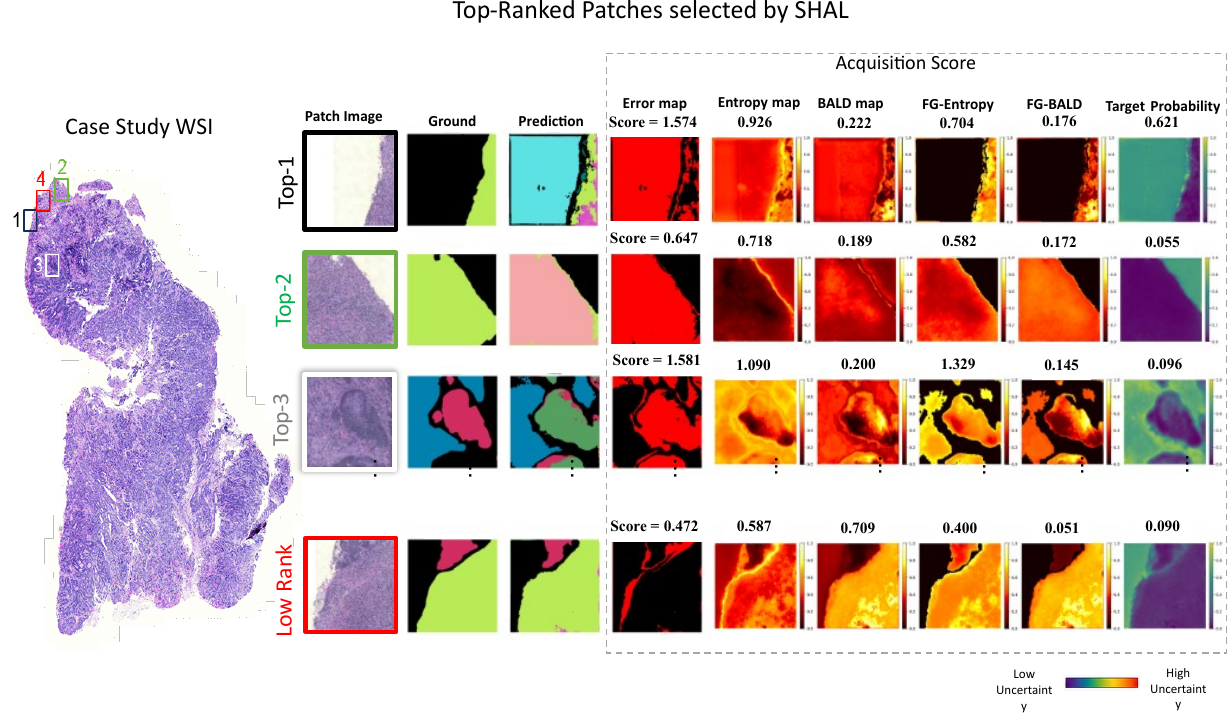}
    \caption{Qualitative example of SHAL acquisition behavior. The selected whole-slide image (WSI) and its top-ranked patches are shown together with the corresponding ground-truth annotations, model predictions, prediction error maps, uncertainty maps, and target-class probabilities. High-scoring patches are consistently associated with morphologically complex tissue structures, elevated uncertainty, and diagnostically relevant target classes, illustrating how SHAL prioritizes informative regions for annotation during active learning.}
    \label{fig:qualitative_results}
\end{figure}

% \begin{figure*}[t]
% \centering
% \includegraphics[width=\textwidth]{Figures/qualitative.png}
% \caption{
% Qualitative comparison of segmentation predictions and uncertainty estimation across different acquisition strategies.
% }
% \label{fig:qualitative_results}
% \end{figure*}
% ###################################################،[ِTODOOOOOOOOOOOOOOOOOOOOOOOOOOOOOOOOOOOOOOOO

\subsection{Annotation Efficiency}

We evaluate annotation efficiency by measuring the number of annotated slides required to reach predefined segmentation performance thresholds on the validation set. Table~\ref{tab:annotation_efficiency} summarizes the annotation budget needed to achieve Dice thresholds ranging from 0.70 to 0.83. Across all performance levels, SHAL consistently requires fewer annotated patients than competing acquisition strategies. At lower performance thresholds ($\geq$0.70 and $\geq$0.75), SHAL reaches the target using only 30 annotated patients, whereas Entropy and Random require 50 patients. This advantage becomes more evident at higher performance thresholds. To achieve Dice~$\geq$~0.80, SHAL requires 50 annotated patients, while all competing methods require at least 70 patients. Similarly, for Dice~$\geq$~0.83, SHAL reaches the target with 70 annotated patients, compared to 90 for BALD and 110 for Random, Entropy, and CoreSet.

These results indicate that SHAL consistently achieves a more favorable trade-off between annotation cost and segmentation performance, demonstrating improved annotation efficiency across a range of performance targets.

\begin{table}[t]
\centering
\caption{Annotation efficiency comparison measured by the number of annotated slides required to reach different Dice score thresholds. Lower values indicate better annotation efficiency.}
\label{tab:annotation_efficiency}
\begin{tabular}{lcccc}
\toprule
\textbf{Method} & \textbf{$\geq$0.70} & \textbf{$\geq$0.75} & \textbf{$\geq$0.80} & \textbf{$\geq$0.83} \\
\midrule
Random         & 50 & 50 & 70 & 110  \\
Entropy        & 50 & 50 & 70 & 110 \\
CoreSet        & 30 & 50 & 70 & 110  \\
BALD   & 30 & 50 & 70 & 90 \\
SHAL (Ours)    & 30 & 30 & 50 & 70  \\
\bottomrule
\end{tabular}
\end{table}

All performance metrics are computed on a fixed test set that remains unchanged across AL rounds, ensuring that improvements reflect the quality of selected training samples rather than variations in the evaluation protocol.

\subsection{Final Segmentation Performance}

Table~\ref{tab:main_results} summarizes segmentation performance across AL rounds on the TCGA test set. All acquisition strategies benefit from additional annotations, with substantial improvements observed between the early and intermediate rounds. At Round 6 (110 annotated patients), SHAL reaches the highest Dice score of 0.846, compared to 0.840 for Entropy, 0.837 for BALD, 0.834 for Random, and 0.827 for CoreSet. This corresponds to an absolute improvement of 0.006 over the strongest baseline (Entropy).
Among the compared methods, BALD exhibits the largest overall improvement from Round~1 to Round~6 (+0.279). SHAL remains competitive throughout the AL process and achieves the best final segmentation performance at the maximum annotation budget. Notably, performance differences among acquisition strategies become smaller in later rounds as additional annotations are incorporated.

\begin{table}[t]
\centering
\caption{Macro Dice score across active learning rounds on the TCGA test set. 
Each round adds 20 annotated patients to the selected pool. Bold values indicate the best performance per round.}
\label{tab:main_results}
\resizebox{\linewidth}{!}{%
\begin{tabular}{lccccccc}
\toprule
\textbf{Method} & 
\textbf{R1} & \textbf{R2} & \textbf{R3} & \textbf{R4} & \textbf{R5} & \textbf{R6} & 
\textbf{$\Delta$ R1$\rightarrow$R6} \\
 & 
\textbf{(10)} & \textbf{(30)} & \textbf{(50)} & \textbf{(70)} & \textbf{(90)} & \textbf{(110)} &  \\
\midrule
Random  & 0.6144 & 0.7327 & 0.7857 & 0.8180 & 0.8279 & 0.8335 & +0.2191 \\
Entropy & 0.6106 & 0.6958 & 0.7511 & 0.8092 & 0.8165 & 0.8399 & +0.2238 \\
BALD    & 0.5581 & \textbf{0.7624} & 0.7884 & 0.8159 & 0.8262 & 0.8370 & \textbf{+0.2789} \\
\rowcolor{gray!10}
\textbf{SHAL (Ours)} & \textbf{0.6161} & 0.6944 & \textbf{0.8143} & \textbf{0.8344} & \textbf{0.8383} & \textbf{0.8457} & +0.2351 \\
CoreSet & 0.5954 & 0.7423 & 0.7781 & 0.8143 & 0.8154 & 0.8274 & +0.2320 \\
\bottomrule
\end{tabular}}
\end{table}

Table~\ref{tab:perclass} reports per-class Dice scores at Round~6. Performance leadership varies across tissue categories, with Entropy achieving the highest Dice scores for tumor and adipose, BALD performing best for stroma, necrosis, and mucosa, and CoreSet obtaining the strongest performance for vessels and muscle. SHAL achieves the highest Dice scores for lymphocytes and benign epithelium while maintaining competitive performance across most remaining tissue classes. Although no single acquisition strategy dominates all categories, SHAL attains the highest overall segmentation performance while maintaining competitive performance across diverse tissue classes.

\begin{table}[t]
\centering
\caption{Per-class Dice at Round~6 (110 annotated patients) on the TCGA
test set. \textbf{Bold} and \underline{underline} denote the best and second-best
score per class. Classes are ordered by prevalence (most to least frequent).}
%Aggregate Macro Dice across rounds is reported in Table~\ref{tab:round6}.}
\label{tab:perclass}
\footnotesize
\setlength{\tabcolsep}{5pt}
\renewcommand{\arraystretch}{1.15}
\resizebox{\linewidth}{!}{%
\begin{tabular}{l *{10}{c}}
\toprule
& \multicolumn{5}{c}{\textbf{Major classes}} & \multicolumn{5}{c}{\textbf{Minority classes}} \\
\cmidrule(lr){2-6} \cmidrule(lr){7-11}
Method & Tumor & Stroma & Benign & Mucosa & Lymph. & Necr. & Adipose & Debris & Muscle & Vessels \\
\midrule
Random      & 0.816 & 0.917 & \underline{0.894} & 0.661 & 0.693 & 0.459 & 0.732 & 0.741 & 0.546 & 0.252 \\
Entropy     & \textbf{0.831} & \underline{0.919} & 0.891 & 0.711 & \underline{0.703} & \underline{0.475} & \textbf{0.823} & \underline{0.753} & 0.553 & 0.372 \\
BALD        & 0.824 & \textbf{0.919} & 0.890 & \textbf{0.745} & 0.696 & \textbf{0.475} & 0.768 & 0.749 & 0.552 & \underline{0.400} \\
CoreSet     & 0.819 & 0.905 & 0.888 & 0.678 & 0.685 & 0.445 & 0.809 & 0.739 & \textbf{0.612} & 0.311 \\
\rowcolor{gray!12}
SHAL (Ours) & 0.811 & 0.916 & \textbf{0.896} & \underline{0.715} & \textbf{0.708} & 0.465 & \underline{0.814} & \textbf{0.757} & \underline{0.567} & \textbf{0.434} \\
\bottomrule
\end{tabular}}
\end{table}

% \begin{table}[t]
% \centering
% \caption{Final segmentation performance on the held-out test set at the maximum annotation budget.}
% \label{tab:final_performance}
% \begin{tabular}{lccc}
% \hline
% Method & Dice $\uparrow$ & mIoU $\uparrow$ & Pixel Acc. $\uparrow$ \\
% \hline
% Random Sampling & $0.615 \pm 0.146$ & $0.455 \pm 0.148$ & $0.565$ \\
% Entropy Sampling & $0.629 \pm 0.139$ & $0.494 \pm 0.137$ & $0.572$ \\
% MC Dropout / BALD & $0.637 \pm 0.131$ & $0.509 \pm 0.128$ & $0.585$ \\
% Hybrid Uncertainty & $\mathbf{0.651 \pm 0.127}$ & $\mathbf{0.510 \pm 0.121}$ & $\mathbf{0.605}$ \\
% \hline
% Full data (100\%) & $0.668$ & $0.523$ & $0.611$ \\
% \hline
% \end{tabular}
% \end{table}

\subsection{Ablation Study}
\label{sec:ablation}

% TODO: insert ablation Dice table here (full / no_stage / no_class / no_fg)

\begin{figure}[htbp]
    \centering
    \includegraphics[width=\textwidth]{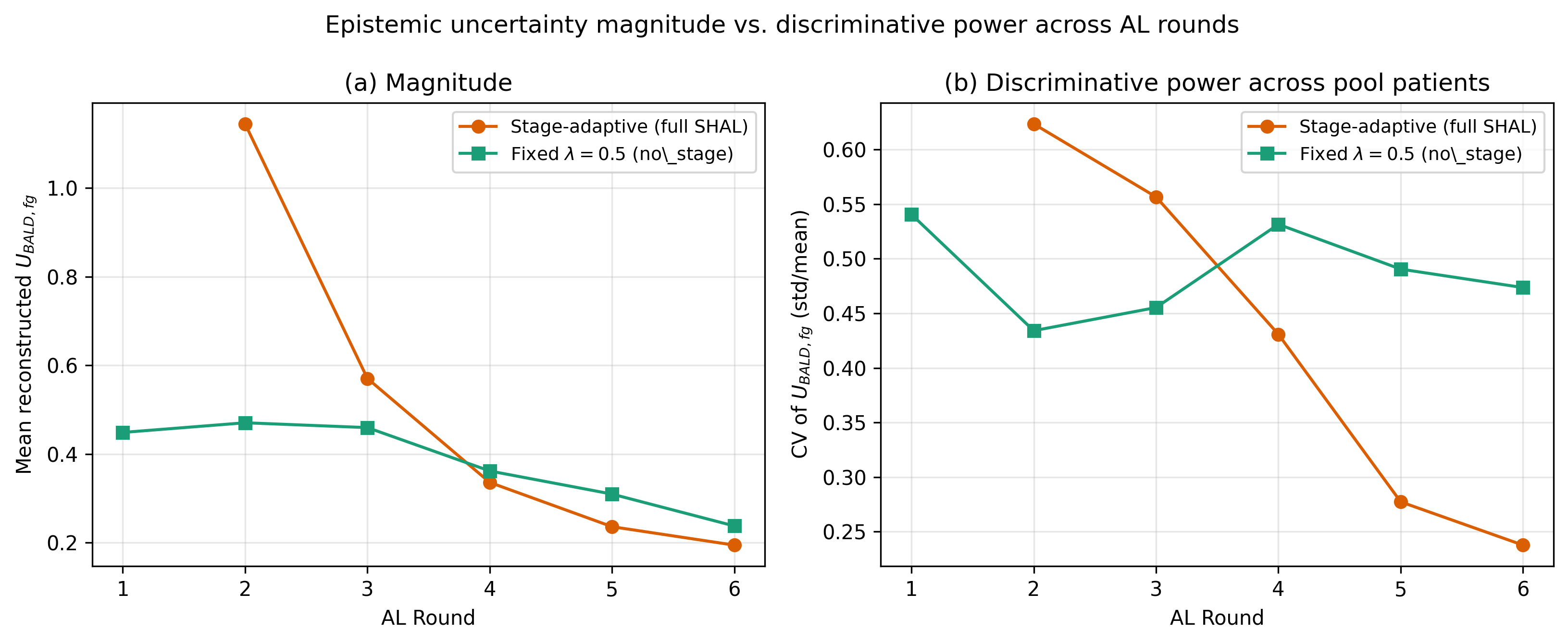}
    \caption{Magnitude and discriminative power of reconstructed epistemic 
    uncertainty ($U_{BALD,fg}$) across active learning rounds, comparing the 
    full stage-adaptive schedule to the \texttt{shal\_no\_stage} ablation 
    (fixed $\lambda=0.5$). (a) Mean $U_{BALD,fg}$ across the unselected pool. 
    (b) Coefficient of variation (std/mean) of $U_{BALD,fg}$ across pool 
    patients, indicating how well the signal discriminates between candidates 
    independent of its absolute magnitude.}
    \label{fig:bald_cv}
\end{figure}

To investigate the mechanism underlying the stage-adaptive schedule, we examined 
how the epistemic uncertainty (BALD) component evolves across AL rounds under 
the full stage-adaptive schedule compared to the \texttt{shal\_no\_stage} 
ablation (fixed $\lambda=0.5$), shown in Figure~\ref{fig:bald_cv}. While the 
mean magnitude of epistemic uncertainty decreases under both schedules 
(Figure~\ref{fig:bald_cv}a), its coefficient of variation across pool patients 
collapses under the stage-adaptive schedule (from 0.62 at Round~2 to 0.24 at 
Round~6) but remains comparatively stable under the fixed-weight ablation 
(0.43--0.54 across all rounds; Figure~\ref{fig:bald_cv}b). This indicates that 
the stage-adaptive schedule's increasing reliance on epistemic uncertainty in 
later rounds coincides with that signal losing most of its ability to 
discriminate between remaining candidates.

To evaluate the contribution of each component of the proposed SHAL 
framework, we conduct an ablation study by systematically removing one 
component at a time while keeping all other settings fixed. We compare 
three variants against the full SHAL model:

\begin{itemize}
    \item \textbf{SHAL-NoFG}: disables foreground-aware uncertainty 
    estimation by setting $w_{fg}(h,w) = 1$ for all pixels, treating 
    background and foreground regions equally during acquisition.
    
    \item \textbf{SHAL-NoStage}: replaces the stage-adaptive weighting 
    schedule with a fixed combination ($\lambda = 0.5$), applying equal 
    weight to predictive entropy and epistemic uncertainty throughout 
    all active learning rounds.
    
    \item \textbf{SHAL-NoClass}: removes the class-aware acquisition 
    bonus by setting $\alpha = 0$, disabling the prioritization of 
    diagnostically relevant minority tissue classes.
\end{itemize}

Table~\ref{tab:ablation} reports the Macro Dice FG across active 
learning rounds for each variant. Removing the stage-adaptive 
component (SHAL-NoStage) leads to the largest performance drop, 
particularly in later rounds where epistemic uncertainty becomes 
increasingly important for identifying informative samples. 
Disabling the foreground-aware weighting (SHAL-NoFG) results in 
degraded performance in early rounds, as uncertainty estimates 
become biased toward uninformative background regions. The 
class-aware bonus (SHAL-NoClass) contributes to improved 
segmentation of minority tissue classes, as reflected in 
per-class Dice scores for underrepresented categories such 
as lymphocytes and vessels.
These results confirm that all three components contribute 
complementarily to the overall performance of SHAL, with the 
stage-adaptive uncertainty combination providing the largest 
individual contribution.

\begin{table}[t]
\centering
\caption{Ablation study: contribution of each SHAL component, measured by
Macro Dice~FG across active learning rounds on the validation set. Each round
adds 20 annotated patients. \textbf{Bold} and \underline{underline} denote the
best and second-best per round.}
\label{tab:ablation}
\footnotesize
\setlength{\tabcolsep}{6pt}
\renewcommand{\arraystretch}{1.15}
\begin{tabular}{l cccccc}
\toprule
Variant & \makecell{R1\\(10)} & \makecell{R2\\(30)} & \makecell{R3\\(50)}
        & \makecell{R4\\(70)} & \makecell{R5\\(90)} & \makecell{R6\\(110)} \\
\midrule
\rowcolor{gray!12}
SHAL (Full)  & 0.309 & \textbf{0.631} & \textbf{0.712} & \textbf{0.703} & \textbf{0.713} & \textbf{0.727} \\
SHAL-NoStage & 0.244 & \underline{0.607} & \underline{0.689} & \underline{0.702} & \underline{0.705} & \underline{0.710} \\
SHAL-NoClass & \textbf{0.404} & 0.628 & 0.651 & 0.686 & 0.700 & \underline{0.722} \\
SHAL-NoFG    & \underline{0.386} & 0.612 & 0.645 & 0.678 & 0.691 & 0.701 \\
\bottomrule
\end{tabular}
\end{table}

\subsection{Generalization to External Cohorts}

We evaluated the generalization of each acquisition strategy on five independent external cohorts (Charité, KAMEDA, LMU, UKK, and WNS) from different clinical centers. As shown in Table~\ref{tab:external_results} %Figure~\ref{fig:external_bar}
SHAL achieves the highest mean external Macro Dice at both Round~3 and Round~6 (0.815). At Round~6, SHAL outperforms Entropy (0.804), BALD (0.807), Random (0.801), and CoreSet (0.790), and ranks first on four of the five external cohorts, with the largest gains observed on UKK and WNS. At Round~3 (50 patients), SHAL attains a mean Macro Dice FG of 0.789, compared to 0.766 for Random, 0.765 for CoreSet, 0.739 for Entropy, and 0.718 for BALD, indicating competitive cross-cohort performance even under limited annotation budgets. SHAL also maintains the smallest internal-to-external generalization gap across all methods at both rounds (0.025 at Round~3 and 0.026 at Round~6), suggesting more consistent generalization to unseen clinical sites.

% ============================================================
% TABLE 2 — External Cohort Generalization @ Round 6
% ============================================================
\begin{table}[t]
\centering
\caption{Generalization performance (Macro Dice~FG) on the internal TCGA test
set and five external cohorts at Round~6 (110 annotated patients).
\textbf{Bold} and \underline{underline} denote the best and second-best score
per cohort. Mean Ext.\ is averaged over the five external cohorts only.}
\label{tab:external_results}
\footnotesize
\setlength{\tabcolsep}{5pt}
\renewcommand{\arraystretch}{1.15}
\resizebox{\linewidth}{!}{%
\begin{tabular}{l ccccc c c}
\toprule
& \multicolumn{5}{c}{\textbf{External cohorts}} & \textbf{Internal} & \\
\cmidrule(lr){2-6} \cmidrule(lr){7-7}
Method & Charit\'{e} & KAMEDA & LMU & UKK & WNS & TCGA & \textbf{Mean Ext.} \\
\midrule
Random  & 0.778 & 0.810 & 0.861 & 0.772 & 0.786 & 0.824 & 0.801 \\
Entropy & 0.784 & 0.811 & \underline{0.869} & 0.770 & \underline{0.786} & 0.827 & 0.804 \\
BALD    & \underline{0.790} & \textbf{0.821} & 0.863 & \underline{0.777} & 0.785 & 0.809 & \underline{0.807} \\
CoreSet & 0.772 & 0.804 & 0.845 & 0.761 & 0.768 & 0.807 & 0.790 \\
\rowcolor{gray!12}
SHAL (Ours) & \textbf{0.796} & \underline{0.815} & \textbf{0.871} & \textbf{0.795} & \textbf{0.797} & \textbf{0.841} & \textbf{0.815} \\
\bottomrule
\end{tabular}}
\end{table}

% \begin{table}[t]
% \centering
% \caption{Generalization performance (Macro Dice FG) on the internal TCGA test set and
%          five external cohorts at Round 6 (110 annotated patients).
%          Bold indicates best performance per cohort.
%          Mean Ext.\ is averaged over the five external cohorts.}
% \label{tab:external_results}
% \resizebox{\linewidth}{!}{
% \begin{tabular}{lcccccc|c}
% \toprule
% Method & Charit\'{e} & KAMEDA & LMU & UKK & WNS & TCGA & Mean \\
% \midrule
% Random          & 0.7775          & 0.8104 & 0.8609          & 0.7716          & 0.7856          & 0.8235          & 0.8012 \\
% Entropy       & 0.7842          & 0.8108 & 0.8694          & 0.7698          & 0.7857          & 0.8274          & 0.8040 \\
% BALD & 0.7900          & \textbf{0.8214} & 0.8627 & 0.7766 & 0.7854          & 0.8094          & 0.8072 \\
% CoreSet       & 0.7724          & 0.8042 & 0.8451          & 0.7613          & 0.7679          & 0.8070          & 0.7902 \\
% SHAL (Ours)        & \textbf{0.7956} & 0.8147 & \textbf{0.8714} & \textbf{0.7952} & \textbf{0.7967} & \textbf{0.8414} & \textbf{0.8147} \\
% \bottomrule
% \end{tabular}}
% \end{table}

Overall, these results suggest that annotation-efficient sample acquisition can contribute to improved cross-domain generalization. This indicates that reducing annotation cost need not come at the expense of performance across independent clinical cohorts.

% \begin{figure}[t]
% \centering
% \includegraphics[
% width=1\linewidth,
% trim={0 1.2cm 0 1.2cm},
% clip
% ]{Figures/generalize.pdf}

% \caption{Generalization evaluation of AL acquisition strategies across internal 
% and external cohorts at Round~3 (50 patients) and Round~6 (110 patients).
% \textbf{(A, C)} Per-cohort Macro Dice FG scores across five external test cohorts 
% (Charit\'{e}, KAMEDA, LMU, UKK, WNS) and the internal TCGA cohort at Round~3 and 
% Round~6, respectively. \textbf{(B, D)} Comparison of internal (TCGA) versus mean external cohort performance 
% at Round~3 and Round~6, respectively. SHAL consistently achieves the highest internal 
% and mean external Dice scores at both rounds (0.814/0.789 at Round~3; 0.841/0.815 at 
% Round~6), while maintaining the smallest generalization gap across all evaluated methods.}
% \label{fig:external_bar}
% \end{figure}

% \begin{figure}[t]
% \centering
% \includegraphics[
% width=1.2\linewidth,
% trim={0 1.9cm 0 1.9cm},
% clip
% ]{Figures/casestudy.pdf}
% \caption{Qualitative example of SHAL acquisition behavior. The selected WSI and its top-ranked patches are shown with segmentation labels, predictions, uncertainty maps, and target-class probabilities. High-scoring patches are associated with complex tissue structures and elevated uncertainty, demonstrating how SHAL identifies informative regions for annotation.}
% \label{fig:slide_selection}
% \end{figure}

\section{Discussion}

% 1. Adaptive uncertainty (چرا hybrid کار می‌کنه)
% 2. Role of random baseline (صادقانه)
% 3. Annotation efficiency و clinical impact
% 4. Generalization و diminishing returns
% 5. Domain shift findings
% 6. Limitations:
%    - Partial annotations
%    - Single architecture tested
%    - Single cancer type (CRC)
% 7. Future work

This study investigates annotation-efficient AL for multi-class histopathology segmentation, with a focus on patient-level slide selection under realistic annotation constraints. The proposed framework demonstrates consistent improvements over baseline strategies across active learning rounds while maintaining strong generalization on multiple external cohorts.

% \subsection*{Acquisition Behavior and Learning Dynamics}

The proposed stage-adaptive acquisition strategy combines predictive entropy 
and epistemic uncertainty through a progressive weighting scheme, transitioning 
from entropy-based exploration in early rounds to dropout-based uncertainty 
in later stages. This progressive mixing proved efficient across the active 
learning process, with the hybrid strategy consistently outperforming static 
single-measure baselines. Predictive entropy performs competitively in early 
rounds, when the model is undertrained and exhibits high ambiguity across the 
unlabeled pool. As training progresses, epistemic uncertainty captured via 
Monte Carlo dropout (BALD) plays a more important role in identifying 
informative samples. We also observe that random sampling performs 
competitively in the earliest rounds, consistent with prior findings in 
AL literature, as uncertainty estimates derived from an 
undertrained model are inherently unreliable. However, once sufficient 
labeled data is available, the proposed method consistently outperforms 
random selection, demonstrating the value of informed sample acquisition.

% A key finding of this work is that different uncertainty measures are effective at different stages of the active learning process. Predictive entropy performs competitively in early rounds, when the model is undertrained and exhibits high ambiguity across a large portion of the unlabeled pool. In this regime, entropy effectively promotes exploration by identifying samples with uncertain predictions. As training progresses, however, entropy becomes less informative. Instead, epistemic uncertainty, captured via Monte Carlo dropout (BALD), plays a more important role in identifying informative samples. This transition reflects the shift from ambiguity-driven exploration to model-driven refinement. The proposed hybrid acquisition strategy explicitly models this behavior by combining entropy and dropout-based uncertainty through a round-dependent weighting scheme. This adaptive design allows the acquisition function to remain aligned with the model’s learning stage, resulting in more consistent performance gains compared to static acquisition strategies. We also observe that random sampling performs competitively in the earliest rounds, consistent with prior findings in active learning literature. This is expected, as uncertainty estimates derived from an undertrained model are inherently unreliable. However, once sufficient labeled data is available, the proposed method consistently outperforms random selection, demonstrating the value of informed sample acquisition.

% \subsection*{Annotation Efficiency and Clinical Impact}

A central contribution of this work is the significant reduction in annotation requirements. The proposed method achieves the target segmentation performance using only 36.48\% of the available labeled data, corresponding to approximately 70 slides instead of 190. In practical terms, this reduction translates into substantial savings in expert annotation time, which is particularly critical in computational pathology, where pixel-level annotation of WSIs is both time-consuming and costly. The results highlight the potential of active learning to enable scalable development of high-quality segmentation models under limited annotation budgets.

% \subsection*{Generalization and Domain Shift}

An important observation is that performance saturates relatively early in the active learning process. Beyond a certain point, additional annotated samples yield only marginal improvements, indicating diminishing returns. This suggests that a carefully selected subset of training data can capture sufficient morphological diversity for robust model learning. This observation is further supported by evaluation on external cohorts. Despite substantial domain shifts across centers, the proposed method maintains consistent performance, indicating that the selected training subset effectively captures diverse and representative tissue patterns. The ability to generalize across variations in staining protocols and scanner characteristics is particularly important for real-world deployment in multi-center clinical settings.

% \subsection*{Limitations}

This study has several limitations. First, the use of partial pixel-level annotations introduces potential bias, as unlabeled tissue regions are treated as background during training. While the proposed foreground-aware uncertainty strategy mitigates this issue, the reported performance should be interpreted within this context. Partially labeled WSIs are however common in practice, due to the costs of generating complete annotations. Segmentation methods should therefore be designed to perform on partially annotated data. Second, the framework is evaluated on a single cancer type (colorectal cancer) and a single segmentation architecture. Although the chosen model represents a strong baseline, the generalizability of the proposed acquisition strategy to other architectures, imaging modalities, and disease types remains to be explored. Third, while the proposed method reduces redundancy at the patch level, it does not explicitly enforce diversity at the slide level. Incorporating diversity-aware selection mechanisms could further improve annotation efficiency and generalization.

% \subsection*{Future Directions}

Future work may explore the integration of semi-supervised and weakly supervised learning to leverage unlabeled tissue regions more effectively. Additionally, iterative annotation strategies, where model predictions guide targeted labeling, may further reduce annotation effort. Incorporating explicit diversity constraints during acquisition and extending the framework to foundation models pretrained on large-scale pathology data also represent promising directions for improving performance and scalability.

\clearpage
% \appendix
% \section*{Appendix}
% \addcontentsline{toc}{section}{Appendix}

%%       \cite{<label>} ==> [1]

%%
%Example citation, See \cite{lamport94}.

\bibliographystyle{elsarticle-num} 
\bibliography{cas-refs}




% %% For numbered reference style

%   \textit{\LaTeX: a document preparation system},
%   Addison Wesley, Massachusetts,

% \end{thebibliography}
\end{document}